# Independent evaluation of state-of-the-art deep networks for mammography

**Osvaldo M. Velarde, Lucas C. Parra**
Biomedical Engineering Department, The City College of New York, New York, NY 10030
<ovelarde@ccny.cuny.edu>, <parra@ccny.cuny.edu>

**Abstract**

Deep neural models have shown remarkable performance in image recognition tasks, whenever large datasets of labeled images are available. The largest datasets in radiology are available for screening mammography. Recent reports, including in high impact journals, document performance of deep models at or above that of trained radiologists. What is not yet known is whether performance of these trained models is robust and replicates across datasets. Here we evaluate performance of five published state-of-the-art models on four publicly available mammography datasets. The limited size of public datasets precludes retraining the model and so we are limited to evaluate those models that have been made available with pre-trained parameters. Where test data was available, we replicated published results. However, the trained models performed poorly on out-of-sample data, except when based on all four standard views of a mammographic exam. We conclude that future progress will depend on a concerted effort to make more diverse and larger mammography datasets publicly available. Meanwhile, results that are not accompanied by a release of trained models for independent validation should be judged cautiously.

**Introduction**

Women in the US have a 13% chance of developing invasive breast cancer at some point in their life [1]. Mortality due to breast cancer can be substantially reduced by early detection [2]. Mammography plays a central role in early detection because it can show changes in the breast well before they are noticeable in a clinical exam [3]. Screening mammography is relatively inexpensive and is thus the standard of care with 40 million mammograms performed each year in the US alone [3]. Detecting cancers in mammography is an image recognition problem that is well suited for computational approaches. Initial efforts with traditional machine learning to aid in diagnosis have been disappointing in clinical practice [4]. With the recent progress in learning of deep neural models there has been a renewed interest in automated breast cancer detection in mammography [5–8]. Due to widespread screening programs, mammography in particular is well-suited for deep-learning models because they perform well whenever large datasets of labeled images are available[9].

By leveraging large datasets from screening mammography, various teams have recently reported remarkable performance at automatically detecting malignant breast cancers using deep-learning [8,10–12]. Unfortunately, these large efforts have not benefited the broader community as the associated datasets are not publicly available. Additionally, many teams do not make the trained models available for independent testing and further development. Furthermore, large computing infrastructures are needed in some instances to train models that have often in excess of 10 million training parameters [8,10–12]. Until such work can be independently reproduced and validated, progress will remain limited. Figure 1 represents the current situation in terms of public release of datasets and models. Depending on the specific objective, validation requires at a minimum a detailed description of the model architecture, ideally specified in open-source code. If training data is not made available, then the pre-trained models parameters are also useful. Smaller test sets are useful for direct comparison with other trained models.

Fortunately some teams have made their model architectures and trained parameters openly available [10,12–16]. Similarly, some teams have released limited mammogram data for research purposes [17–19]. The goal of the present work is to validate the available state-of-the-art models on these publicly available datasets, and to establish reproducibility and generalization. Note that none of the datasets current available come close to the size of the databases required to train most deep networks. Thus, here we can only evaluate existing fully-trained models. With the somewhat larger datasets, we also attempt to fine-tune the model. Additionally, we tested a few approaches to harmonize data in image size, resolution and intensity distribution.

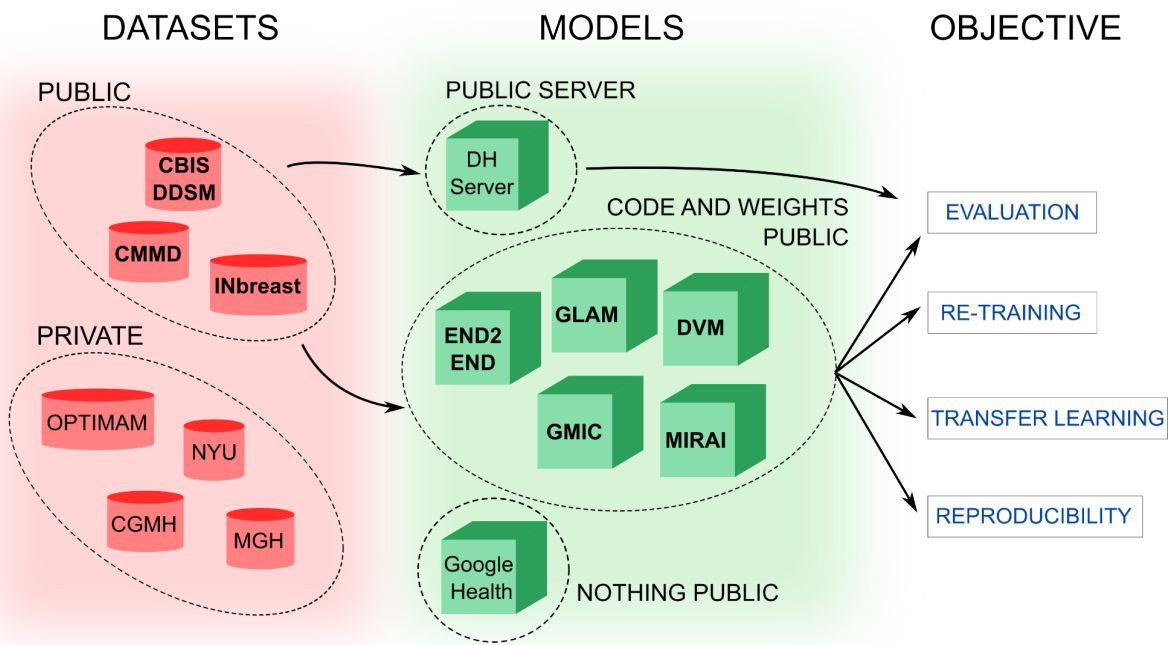

**Figure 1: Current availability of mammography datasets and deep-learning models.** The datasets and models listed are for publications [10,12–19]. Those marked in bold were available in this study and are listed in Table 1 and Table 2, respectively. Different research objectives include: Evaluation: A model has been trained on a large dataset A and one would like to know how well it performs on dataset B, that is perhaps too small to retrain the model. For this, one requires both the model architecture (in code) as well as the model weights. Re-training: When the new dataset B is large it may suffice to have the model architecture and one can retrain the model from scratch. Transfer learning: A trained model can be used as a feature extractor (e.g. segmentation network) to use for detection. In this case code and trained weights need to be available. Reproducibility: One may want to reproduce the published results on the original data and model. For this the model architecture needs to be available as well as the original dataset A.

**Results**

Published models have largely been trained and evaluated using private databases. Here we evaluated the ability to generalize to an out-of-sample dataset. To this end, we compiled three publicly available datasets: CBIS-DDSM, INbreast, CMMD (see Section Methods: Datasets). We evaluated models available to us through public release of code (see Section Methods: Models). This includes three models from the group of Krzysztof Geras at NYU (GLAM, GMIC, DVM), a model from the group of Regina Garzilay at MIT (MIRAI) and a model from the group of Li Shen at MSSM (End2End). Another model available is a product of Deep Health, Inc (DH) that is a server to obtain the predictions of its model remotely (i.e., the data must be sent to the server). However, they only accept images obtained by a Hologic system. Since the datasets used do not meet this requirement, we were unable to evaluate their performance.

The resulting ROC curves for all combinations of models and data is shown in Fig. 2. The corresponding area under the curve as summarized in Table 1, which also lists performance as reported in the original publications, on private datasets. The general observation is that the published results are all numerically higher than the performance of the same models on the public datasets (statistical comparison is not feasible as we do not have full ROC curve from these published results). There are only two exceptions to this. The End2End model was tested originally on a subset of the public CBIS-DDSM data. There we reproduce the results exactly. However, on the full CBIS-DDSM test data, performance is numerically lower. The other exception is the DVM model, which seems to generalize well to the CMMD and INbreast data with performance numbers comparable to the previously published results (0.9, 0.85 vs the previous 0.88). It should be noted that the MIRAI model was not designed for diagnosis but rather for risk prediction. Its objective is to take a mammogram from "today" to predict cancer status in subsequent years. Here risk prediction for year 0 was used to predict cancer diagnosis "today", and so the lower performance observed should be viewed with caution. In summary, only DVM seems to generalize well to public datasets, which may be due to the fact that it leverages all four standard views of a typical exam, i.e. bilateral craniocaudal (CC) and mediolateral oblique (MLO) views for both breast. The CBIS-DDSM dataset seems to be challenging for all models. A possible cause for this may be that the distribution of image sizes is much wider for the CBIS-DDSM data as compared to the INbreast data (Fig. 7b). A similar behavior is observed in the intensity distribution (Fig. 7a).

|  |  | Model | | | | |
|---|---|---|---|---|---|---|
|  |  | **E2E** | **GLAM** | **GMIC** | **DVM** | **MIRAI** |
| Properties | Group | **MSSM** | NYU | NYU | **NYU** | MIT |
|  | Code | Yes | Yes | Yes | Yes | Yes |
|  | Input | single image: CC or MLO view of one breast | | | 4 views: left/right CC, MLO | |
| Datasets evaluated here | CBIS-DDSM (test) | 0.7 | 0.5 | 0.51 | 0.54 | 0.56 | 0.5 |
|  | INbreast | 0.67 | 0.7 | 0.84 | 0.75 | 0.85 | 0.76 |
|  | CMMD | 0.53 | 0.76 | 0.80 | 0.79 | 0.9 | 0.51 |
| Published previously | Performance (data, year) | 0.85 (CBIS-DDSM*, 2019) | 0.82 (NYU, 2021) | 0.91 (NYU, 2020) | NA | 0.88 (NYU, 2020) | 0.9 (CGMH, 2021) |

**Table 1: Area under the ROC (AUC) on the classification for various published models and datasets.** For the End2End, GLAM and GMIC models, we show the performance on the image-level classification; while for DVM, we show the breast-level classification performance. Finally, for MIRAI we show the exam-level classification performance. See Methods-Metrics section for the definition of AUC on image/breast/exam-level classification.

The results presented in Table 1 and Fig. 2 show the performance of the models for image-level predictions (see Methods-Metrics). In Stadnick et al[20], similar results are shown for breast-level predictions. We were able to independently reproduce Table 2 of Stadnick et al and verify the performance of the algorithms (data not shown).

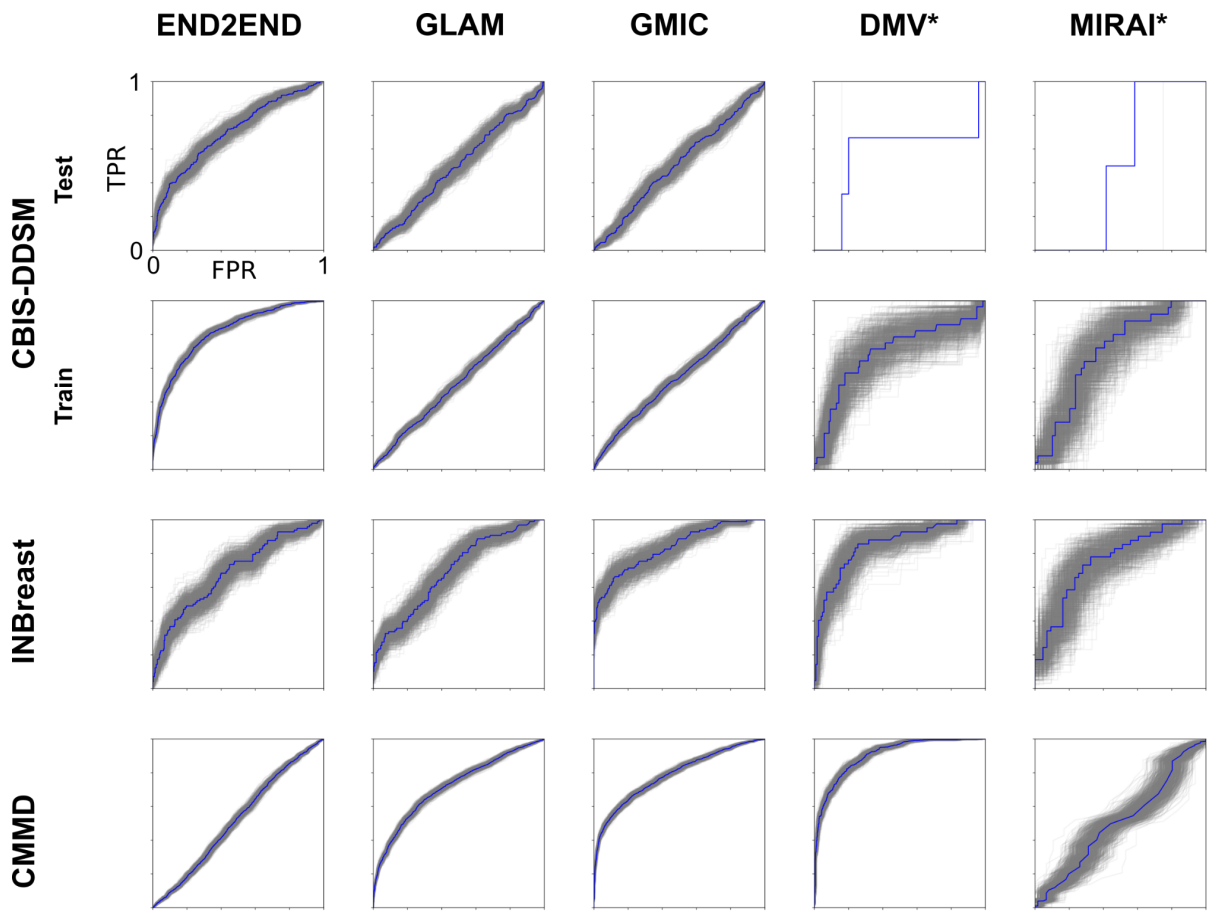

**Figure 2: ROC for various published models and datasets.** For each model, the prediction was calculated for all the images of the corresponding dataset (image-level predictions). With the set of predictions and labels, the ROC curve was determined on the TPR vs. FPR axes (blue curves).

Then, the bootstrap procedure was used to resample with replacement the pairs $(y, \hat{y})$ several times. For these new sampled sets, the ROC curve was calculated (gray curves). These curves are useful for the determination of the area under the curve (AUC) and their confidence interval. The number of bootstrap replicates defines the variance of the estimate. In most cases, we set this value to 1000. Models indicated with * are those that need a full review (i.e, four standard views). For these models, the number of bootstrap replicates was set to 5 when considering the CBIS-DDSM test dataset. This is because the number of tests with four views and label 1 was very small.

The above results used the images without any additional harmonization to the preprocessing of the algorithms to crop, scale, or adjust the images. Fig. 3 shows samples of preprocessed mammograms and the prediction of different models for each one. For all these cases, the label was 1. The images in the upper panel are those whose prediction was the closest to the label (i.e., $min\ |\hat{y} - y|$ ), while those in the lower panel are those whose prediction was the furthest from the label (i.e., $max\ |\hat{y} - y|$ ).

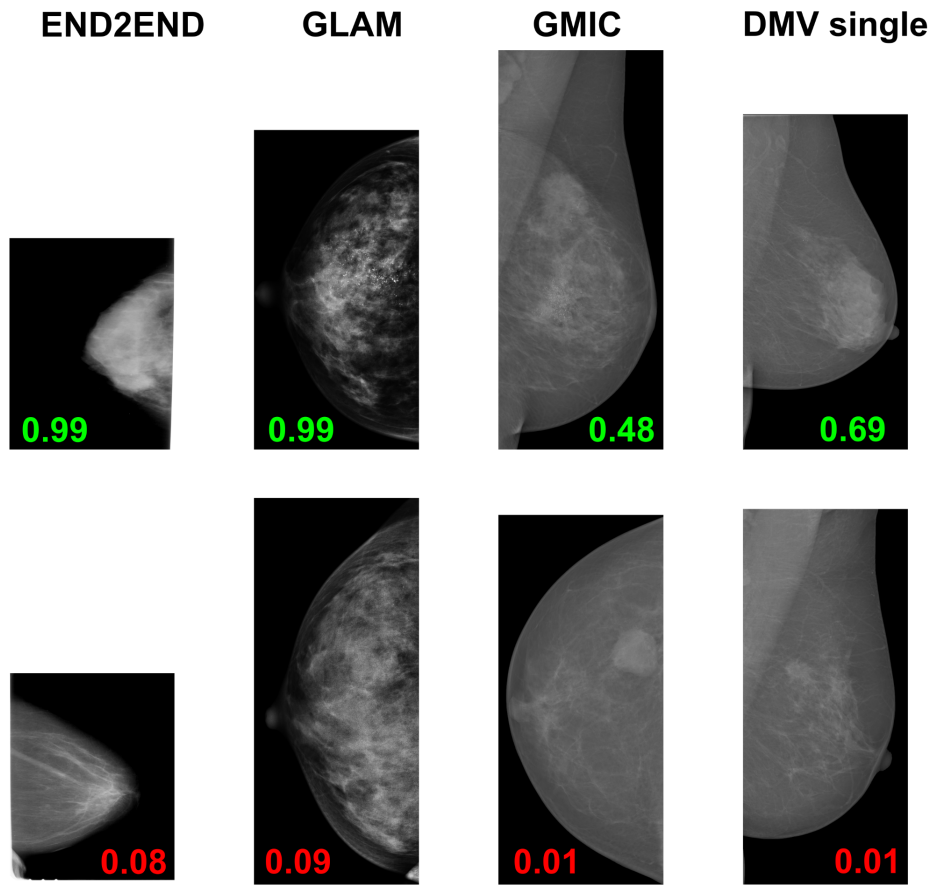

**Figure 3: Examples of preprocessed images and the predictions determined by different models.** In all cases, the label of the images was $y = 1$. In the upper panel, the images correspond to those that the prediction is closest to the label (i.e., $min\ |\hat{y} - 1|$). While in the lower panel, they are those that the prediction is farthest from the label (i.e., $max\ |\hat{y} - 1|$).

Similarly, thus far no effort was made to retrain or fine-tune the models to adjust to differing statistics in the images. Indeed, both image size, resolution, and intensity distributions differed across images (see Fig. 7). We therefore next consider harmonization and retraining. In Fig. 4a, we show the result of applying different intensity harmonization methods on an image and its corresponding cumulative density function (CDF) for intensity [21–23]. We calculate the performance of the DVM model on each dataset when harmonization method is applied to the images (see Fig. 4b). Also, Fig. 5 shows the effect of considering intensity harmonization on the performance of the DVM model in the fine-tuning stage. We observed that the harmonization does not substantially modify the performance, if anything, it reduces performance. In this analysis, we use harmonization operations common in image processing such as equalization and matching.

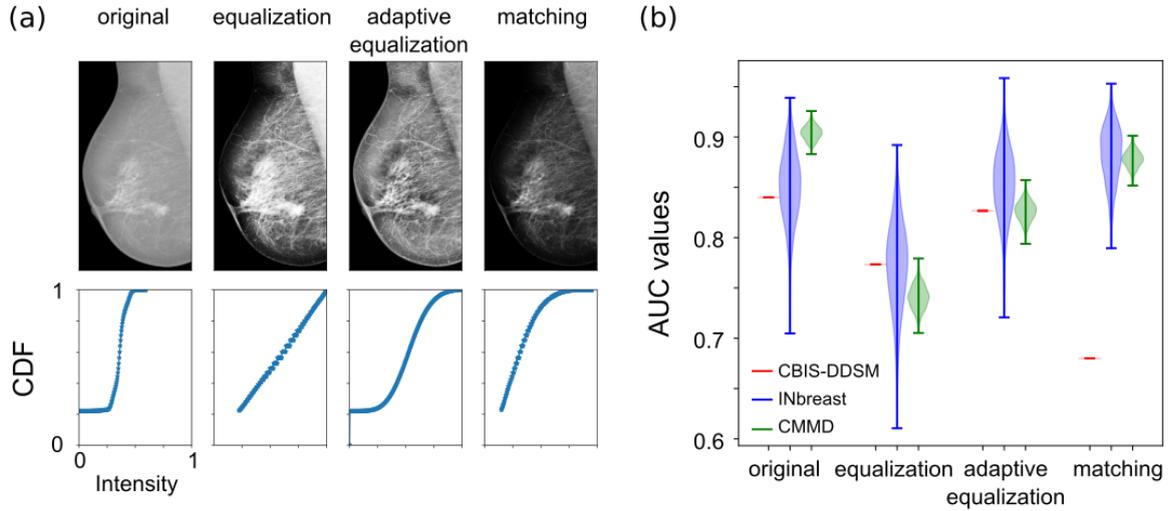

**Figure 4: Effect of intensity harmonization (a)** Cumulative density function (CDF) of images for intensity harmonization: For one image from each dataset, we compute CDF. In addition, we transform it using different intensity harmonization methods (equalization, adaptive equalization and histogram matching). We show the CDF for the image resulting from each transformation. **(b)** Evaluation of DVM considering different intensity harmonizations: For the different datasets (red: CBIS-DDSM, blue: INbreast, green: CMMD), the predictions corresponding to the images were calculated considering an additional intensity harmonization operation (equalization, adaptive equalization and histogram matching) to the preprocessing of the algorithm. For histogram matching, we transform each image so that its histogram matches the histogram of a β-distribution with parameters $\alpha = 2$, $\beta = 5$. Note that the only case where performance increases is when considering histogram matching on the INbreast dataset.

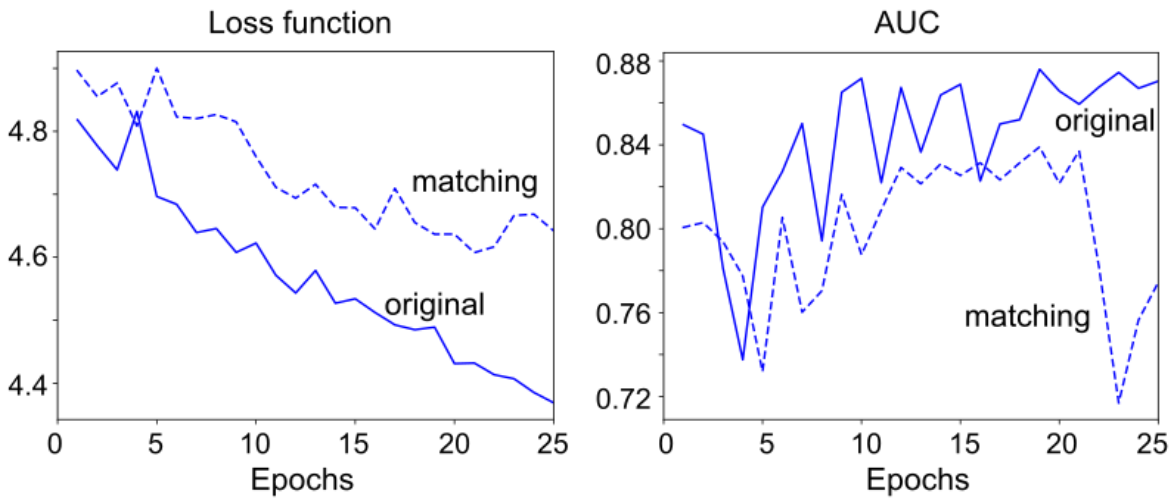

**Figure 5: Fine-tuning of DVM using INbreast dataset.** For the training, the original images of the dataset (solid curves) and the images under the histogram matching operation (dashed lines) were considered. The same hyperparameters reported in Wu et al [15] were considered for the fine-tuning of the last layers of DVM. The weights of the convolutional and batch normalization layers were kept fixed. As in the evaluation stage, the model's performance does not improve with intensity harmonization.

**Discussion**

Recently a number of reports documented impressive performance of deep-learning models in mammography diagnosis, at times matching performance of trained radiologists [10–16]. However, much of what has been published is difficult or impossible to reproduce, for a variety of reasons; e.g, the availability of the code, the availability of training data, or a lack of details about the training process. In this work we made a concerted effort to independently evaluate all publicly available models on public datasets. Given the limited size of publicly available datasets, in this work we only considered pre-trained models for evaluation and fine-tuning.

Classification performance was evaluated here in terms of AUC for individual mammographic views, as well as combining views of a given breast. These "breast-level" predictions reproduce the results previously published for all model tests in Stadnick et al[20]. This confirms that the published models are complete and that we properly implemented their use, which is important in the context of new data.

On new data, models generally underperformed when using individual views. For the CBIS-DDSM dataset, performance had previously been reported on a selected subset of the data. When performance is evaluated on the complete set, all models dropped significantly in performance, suggesting that these data are more challenging.

The models tested here differ in the way they integrate global and local image information as well as the ways of combining mammography views to predict benign and malignant lesions. A model that stands out with good generalization performance is the DVM model from NYU[15]. We attribute this to the combination of the two mammographic views from both breasts. In our view, this pre-trained model is a good starting point for further research in algorithm development.

We also analyzed the effect of an additional pre-processing of the data on both the performance and the fine-tuning process. We used harmonization methods generally in line with previous efforts [21–23]. Generally we find that preprocessing aimed at harmonizing the data did not improve performance. In fact, performance was very sensitive to any kind of image manipulation, and therefore we believe it is difficult to fine tune harmonization for existing pretrained models.

With the advancement of deep learning, the medical imaging community is interested in applying these techniques to improve the accuracy of cancer screening and the development of a new generation of Computer-Aided Detection tools[24]. In addition to diagnosis, deep learning models have been developed for risk prediction and lesion segmentation. Although these tasks are distinct, transfer learning techniques can be used to leverage training data across these complementary tasks. For example, the use of detailed cancer region annotations (data usually used for segmentation) could be used to improve image-based risk models [15,25]. A crucial point is that deep learning models' generalization ability relies on the structure of the model and its training (i.e., algorithm, size and quality of the data available for training). Our results suggest that generalization will require large but also diverse datasets with high-quality labels for training. The alternative may be to use a narrowly defined acquisition protocol and scanner type, which is the approach taken from some commercial efforts[12]. Our overall conclusion is that automatic diagnosis of mammography should be regarded with care when it is used outside of the context in which it was developed. Furthermore, the field can only make progress as a whole if models are shared with code and pre-trained parameters and if larger datasets become publicly available. In short, without replication there will be no science of AI for Radiology.

## Methods

### I- Datasets

In the following sections we indicate some characteristics of each dataset. Table 2 shows information about each dataset. On the other hand, Fig. 7 shows the density function of the pixel intensity and the distribution of image sizes for each dataset. The following subsections provide some details for each dataset.

| Dataset name | Intensity (8 bits) | | Database size | | |
|---|---|---|---|---|---|
| | Mean | Standard deviation | Num of exams | Num of images | Num of exams (with 4 views) |
| CBIS-DDSM | 53 | 56 | 1597 | 3130 | 98 |
| INbreast | 30 | 31 | 115 | 410 | 84 |
| CMMD | 18 | 65 | 1775 | 5202 | 826 |

**Table 2: Information about the datasets used in this work.**

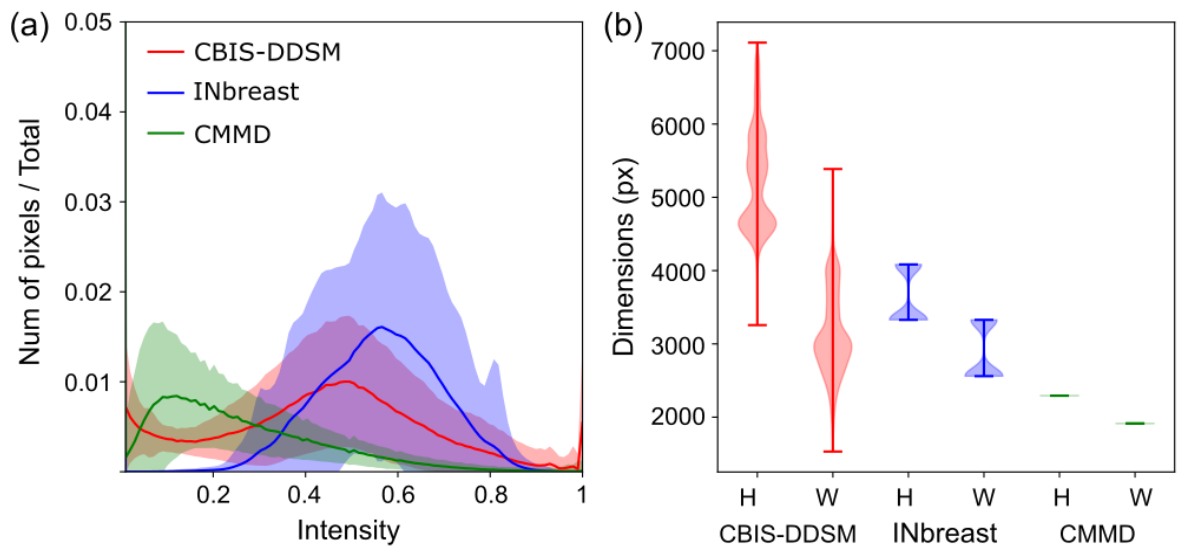

**Figure 7: Intensity histograms and image sizes for all datasets used.** A dataset has $N$ images $\{X_j\}_{j=1}^{N}$ of different 2D-spatial dimensions $\{(H_j, W_j)\}_{j=1}^{N}$. For each image, the pixel intensity histogram $P_j(I)$ was calculated (i.e, the fraction of pixels with intensity $I$). **(a)** Each solid line represents the average histogram $<P_j(I)>_j$ for a dataset . The shaded areas indicate the standard errors of the histograms. **(b)** The violin plots show the spatial dimension distributions ($\{H_j\}_{j=1}^{N}$: Height or $\{W_j\}_{j=1}^{N}$: Width) for each dataset.

### I-1. CBIS-DDSM

Digital Database for Screening Mammography (DDSM) is the result of the collaborative effort between the Massachusetts General Hospital, the University of South Florida, Sandia National Laboratories, Washington University School of Medicine, Wake Forest University School of Medicine, Sacred Heart Hospital and ISMD, Incorporated. DDSM is a database of 2620 exams and 10480 images of scanned film mammography studies. It contains normal, benign, and malignant cases with verified pathology information.

With the aim of solving different challenges presented by the original DDSM version for the evaluation of CAD systems research in mammography, Lee et al.[18] developed an updated and standardized version of DDSM. Curated Breast Imaging Subset of DDSM (CBIS-DDSM) collection includes a subset of the DDSM data selected and curated by a trained mammographer. An example of an update is that ROI segmentation and bounding boxes, and pathologic diagnosis for training data are also included. For details of the processing it is recommended to read Lee et al[18].

The dataset can be downloaded from TCIA website [26]. For this dataset there is a standard separation of the training and test subsets. Test set consists of 349 exams (645 images), while the training set consists of 1248 exams (2458 images). As indicated in Table 2, only 6 % of exams have the four views available.

### I-2. INbreast

The INbreast database is a mammographic database whose images were collected at the Breast Centre of the Hospital de São João in Porto, Portugal, between April 2008 and July 2010. It has a total of 115 exams (410 images) of which 73 % cases are from women with both breasts (4 images per case) [17].

The database includes examples of normal mammograms, mammograms with masses, mammograms with calcifications, architectural distortions, asymmetries, and images with multiple findings. The size of the images are 3328 x 4084 or 2560 x 3328 pixels.

INbreast dataset was released in 2012 and can be downloaded from Kaggle website [27]. There is no standard division of test and training subsets for INbreast. In this work, the dataset was not separated into subsets.

### I-3. CMMD

The Chinese Mammography Database (CMMD) was collected between 2012 and 2016, in Sun Yat-sen University Cancer Center and Nanhai Hospital of Southern Medical University in Foshan. In 2021, it was published by The Cancer Imaging Archive (TCIA) and it can be downloaded from TCIA website [28].

The database CMMD consists of 1775 patients/exams (5202 images): (1) 3728 mammographies with biopsy confirmed type of benign or malignant tumors (2) 1498 mammographies with additional information about molecular subtypes.

Dataset images are accompanied by biopsy-proven breast-level benign and malignant labels. Dataset authors also provided age and finding type (calcification, mass or both) for all patients as well as immunohistochemical markers for 749 patients with invasive carcinoma.

### II - Models

An image is represented by a tensor $X$ of the form $H_0 \times W_0 \times Ch_0$ where $H_0, W_0$ indicate the height and width of the image, and $Ch_0$ indicates the number of channels (usually 1 or 3). A convolution is an operation of the form

$$Y[p,q,c] = \sum_{i,j,k} F[p-i, q-j, c, k]\, X[i,j,k]$$

where the tensor $F$ is called the *kernel/filter of the convolution*. In addition to the dimensions of $F$, two additional parameters are usually specified: padding ($p$) and stride ($s$). The operation of *padding* is to add zero value pixels around the image $X$. On the other hand, the kernel shifts a certain number of pixels to the right/down, during each iteration of convolution. This offset is known as *stride*. One of the main advantages of these parameters is to allow modifying the dimensions of the output.

The result of the convolution is a new tensor $Y$ with new dimensions $H_1 \times W_1 \times Ch_1$. This output is known as the feature map of $X$. In Fig. 8, the relation between the dimensions of the input and the output of a convolution is shown.

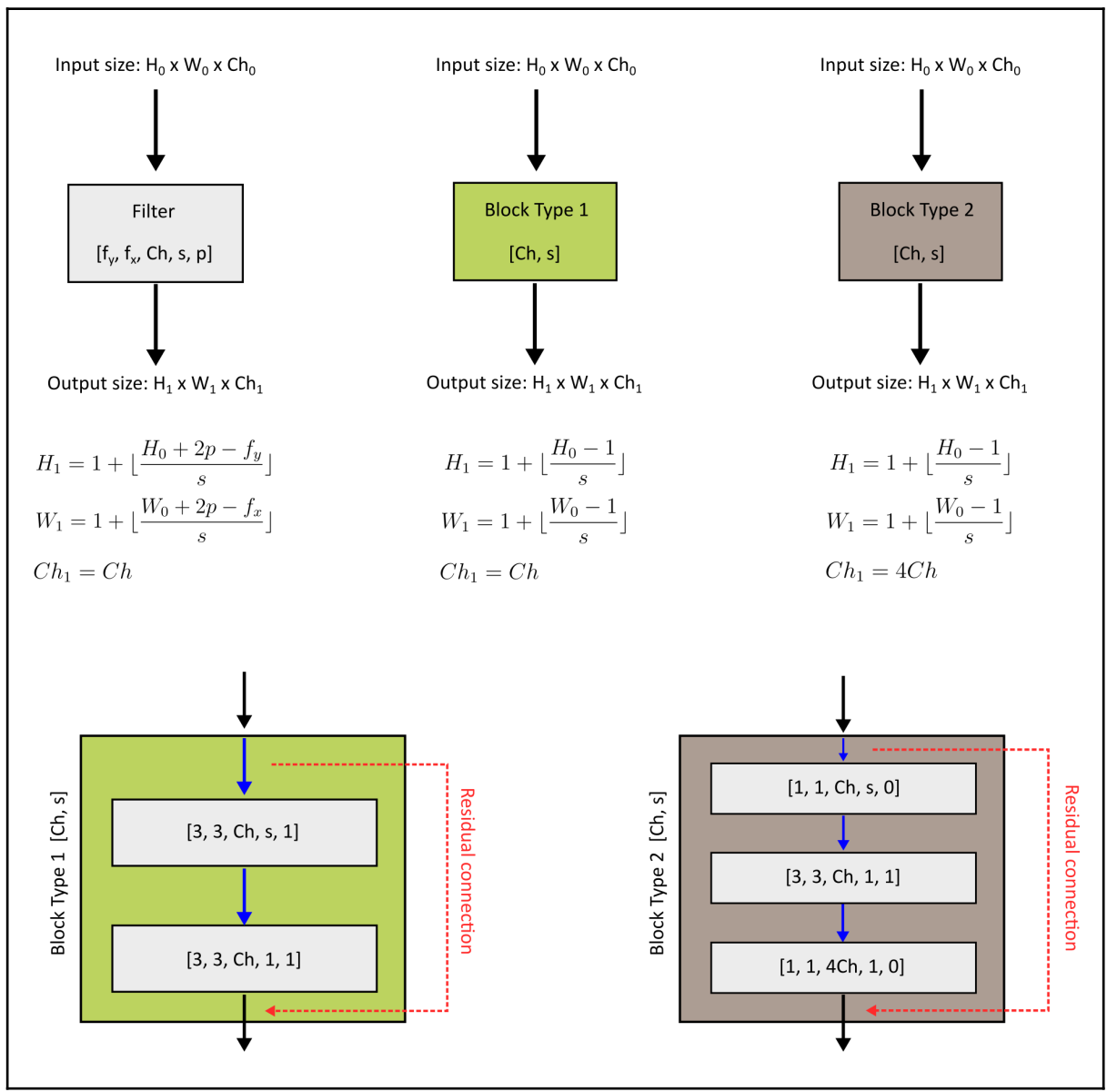

**Figure 8: Structure of the blocks usually used in convolutional networks.** A convolution is characterized by the filter/kernel with spatial dimensions, channels, and padding and stride parameters. The dimensions of the output of the operation depend on the dimensions of the input and the kernel parameters. Blocks (types 1, 2) are ways to group convolutions sequentially. Sometimes these blocks consider a residual connection that connects the input to the output directly. The blue arrows inside the blocks represent extra operations that do not modify the dimensions of the tensor (e.g, batch normalization or non-linear function)

One of the most relevant constructions of Machine Learning in recent years is the concept of Deep Convolutional Network. This idea is inspired by the anatomy and functionality of the visual system. In fractions of seconds, humans can identify objects within our field of vision. They can also name these objects, perceive their depth, perfectly distinguish their outlines, and separate them from their backgrounds. Somehow the system captures pixel data and transforms that information into more meaningful features; e.g, geometry, abstractions, semantic meaning.

In the neuroscientific literature, it is argued that the visual cortex contains a complex hierarchical arrangement of neurons [29]. Visual information is introduced into the cortex through the primary visual area (V1) that deals with low-level visual features (e.g., contour segments, contrast, and color). Then, V1 transmits this information to other areas (V2, V4 and V5) that deal with the more specific or abstract aspects of the information. Based on this, the image processing community tends to group different kernels sequentially to form more complex blocks or structures such as deep convolutional neural networks.

In Fig. 8, two ways of grouping a pair of filters called Block Type 1 and Type 2 are shown. These blocks are the minimum basic structure for the formation of convolutional networks known as *Resnet*. Note that these two types of blocks only have two parameters $[Ch, s]$. For example, the notation block type 1 $[Ch = 5, s = 2]$ means that you are considering two consecutive filters: (1) $3 \times 3 \times 5$ - filter with $s = 2$, $p = 1$ and (2) $3 \times 3 \times 5$ - filter with $s = 1$, $p = 1$. Note that as before, the blocks modify the input dimensions based on the $[Ch, s]$ parameters (see equations of Block Type 1,2. in Fig. 8).

Usually, the difficulty of training a deep network increases with the number of layers, due to problems such as exploding gradients and vanishing gradients. Residual connection is a simple and very effective technique to improve training deep neural networks in terms of convergence. It provides another path for data to reach latter parts of the feedforward neural network by skipping some layers. The residual connection applies a mapping $\psi$ to the input $X$, then it performs element-wise addition $\phi(X) + \psi(X)$ (where $\phi(X)$ represents the output of the feedforward connections). Usually, $\psi$ is the identity function; however, in the models considered in this work, the residual connection $\psi$ is a filter $[1, 1, Ch, s, 0]$. In Fig.8, we indicate the possible residual connections in Block Type 1,2. We denote R-Block Type 1,2 to the case that the Block Type 1,2 considers a residual connection.

As mentioned above, convolutional networks are the result of a sequential grouping of different kernels. Basically, convolutional neural networks are very similar to ordinary neural networks like multilayer perceptrons. Generally, they are built of convolutional layers (convolutions/filters or blocks), reduction layers (pooling) and a fully connected classifier layer, which will give us the final result of the network (prediction). An example of convolutional networks is the Resnet type. Table 3 shows the structure of some Resnet networks. Each column corresponds to the architecture of a type of Resnet. Note that each network is a combination of 5 or 6 stages and each stage is a set of blocks. For example, Resnet-18 is a combination of stage 0, I, II, II, IV. For this case, stage 0 is a combination of a filter and a max pool operation. In this same network, stage IV is the succession of two R-Blocks Type 1 with $Ch = 512$ and $s = 2$.

| Stage | 18-layers | 34-layers | 50-layers | 22-layers |
|---|---|---|---|---|
| 0 | Filter [3,3,64,2,3] Max Pool [3,3,64,2,3] | Filter [3,3,64,2,3] Max Pool [3,3,64,2,3] | Filter [3,3,64,2,3] Max Pool [3,3,64,2,3] | Filter [3,3,16,2,3] Max Pool [3,3,16,2,3] |
| I | R-Block Type 1 [64,1] x 2 | R-Block Type 1 [64,1] x 3 | R-Block Type 2 [64,1] x 3 | R-Block Type 1 [16,1] x 2 |
| II | R-Block Type 1 [128,2] R-Block Type 1 [128,1] | R-Block Type 1 [128,2] R-Block Type 1 [128,1] x 3 | R-Block Type 2 [128,2] R-Block Type 2 [128,1] x 3 | R-Block Type 1 [32,2] R-Block Type 1 [32,1] |
| III | R-Block Type 1 [256,2] R-Block Type 1 [256,1] | R-Block Type 1 [256,2] R-Block Type 1 [256,1] x 5 | R-Block Type 2 [256,2] R-Block Type 2 [256,1] x 5 | R-Block Type 1 [64,2] R-Block Type 1 [64,1] |
| IV | R-Block Type 1 [512,2] R-Block Type 1 [512,1] | R-Block Type 1 [512,2] R-Block Type 1 [512,1] x 2 | R-Block Type 2 [512,2] R-Block Type 2 [512,1] x 2 | R-Block Type 1 [128,2] R-Block Type 1 [128,1] |
| V | - | - | - | R-Block Type 1 [256,2] R-Block Type 1 [256,1] |

**Table 3: Structure of the ResNets involved in the models considered in this work.** The notation is described in the text.

In the supervised learning problem, three elements are considered: (1) a training data set $D = \{x_n, y_n\}$, (2) a proposed model $F(x, \theta)$ with trainable parameters $\theta$, and (3) a cost function $L(D, F)$ to minimize.

In the case of the models studied, the input $x_n$ is a mammogram or an exam. An exam consists of a collection of mammograms of both breasts (left/right) in different views (MLO / CC). The output of the models $\hat{y}$ can be:

1. a binary type variable and $\hat{y} \in [0, 1]$ for the classification of a mammogram (cancer/normal). The End2End model considers this type of output.
2. a two-dimensional vector $\hat{y} \in [0, 1]^2$ where each component is an estimation of the probability of the presence of malignant and benign lesions (i.e, $\hat{y} = [y_m, y_b]$). This is the approach implemented in GLAM and GMIC models.

3. a four-dimensional vector $\hat{y} \in [0, 1]^4$ where each component is an estimation of the probability of the presence of malignant and benign lesions for each breast (i.e, $\hat{y} = [y_{mL}, y_{mR}, y_{bL}, y_{bR}]$).

4. a five-dimensional vector $\hat{y} \in [0, 1]^5$ where each component is an estimation of the cancer risk for each year over the next 5 year (i.e, $\hat{y} = [risk_{1\,year}, risk_{2\,year}, risk_{3\,year}, risk_{4\,year}, risk_{5\,year}]$).

Regarding the general structure of the models, there are notable differences between them. The End2End model only consists of a convolutional network (see Table 4). DVM model consists of 4 convolutional networks working in parallel (one CNN per view), then the outputs are combined.

The GLAM and GMIC models consider three modules. The global module is a memory-efficient CNN that extracts the global context and generates prominence maps that provide an approximate location of possible benign/malignant findings. On the other hand, the local module is a CNN with greater capacity to extract detailed visual details of regions of interest (patches) and then condensing the information using some attention or aggregation mechanism. Finally, the fusion module combines the representation vectors of the global and local modules to produce a fusion prediction. The main difference between GLAM and GMIC lies in the global module. On the one hand, in GLAM, the global module provides a pyramidal hierarchy of multi-scale features maps when processing an input image, whereas GMIC only provides a single-scale feature map.

In the MIRAI model, each mammogram view is processed independently through the image encoder. Next, an aggregation module combines information across views and obtains a representation of the entire mammogram using attention-pooling. Finally, the model can predict risk factor information and a patient's risk with an additive-hazard layer.

| Model | CNN | Num of parameters | Input | Output | Loss function | Training dataset |
|---|---|---|---|---|---|---|
| End2End (MSSM) | Resnet-50 VGG16 | 23 M 134 M | One single mammography image | Probability of malignancy in the breast. | BCE between output and label | CBIS-DDSM INbreast |
| GLAM (NYU) | Resnet-22 (global network) Resnet-34 (local network) | 21 M | | Probability of: 1- malignancy. 2- benign lesion. | Sum of BCEs obtained in the different modules (global, local and fusion). | NYU Breast Cancer Screening Dataset (186,816 exams with four views) |
| GMIC (NYU) | Resnet-22 (global network) Resnet-18 (local network) | 11 M | | | | |
| DVM (NYU) | Resnet-22 | 21 M | One exam with 4 views: | Probability of: 1- left breast malignancy. 2- right breast | Sum of BCEs. | |

| | | | R-CC, L-CC, R-MLO, L-MLO | malignancy. 3- left breast benign lesion. 4- right breast benign lesion. | | |
|---|---|---|---|---|---|---|
| MIRAI (MIT) | Resnet-18 | 11 M | | Risk for each year over the next 5 year | Log-likehood | Massachusetts General Hospital (MGH) Dataset (210,819 exams with four views) |

**Table 4: Information about the models analyzed in this work.**

### III - Metrics

Consider a set of exams $\{E_i\}$ where each exam consists of one or more images, i.e. $E_i = \{X^{(i,s,v)}\}$. In this notation, $i$, $s$ and $v$ indicate the index of exam, laterality (s = LEFT, RIGHT) and view of the breast taken during mammography (v = MLO, CC), respectively.

For these exams, there is a set of binary labels $\{y_m^{(i,s)}\}$ that indicate whether the breast $s$ of exam $i$ presents a malignant lesion. On other hand, the End2End, GLAM and GMIC models return the probability of finding a malignant lesion $\hat{y}_m^{(i,s,v)}$ for each mammogram $X^{(i,s,v)}$ (see Models).

Both the predictions and the labels are necessary for the calculation of the model's performance (i.e, AUC). Two types of predictions are defined:

1. image-level predictions: $\hat{y}_m^{(i,s,v)}$
2. breast-level predictions: $\hat{y}_m^{(i,s)} = \frac{1}{|V|}\sum_v \hat{y}_m^{(i,s,v)}$

where V is the number of available images with view $v$ for the breast $s$ of the exam $i$. AUC on the image/breast-level is defined as AUC considering the image/breast-level predictions.

It is important to note that the DVM model only returns breast-level predictions (see Models), while for End2End, GMIC and GLAM it is possible to calculate both types of predictions.

On the other hand, in our evaluation of MIRAI we consider that the estimate of the cancer risk for next year ($risk_{1year}$) is an approximation of the diagnostic task. In this case, the model only generates a single prediction for the entire exam (exam-level prediction).

### Acknowledgements
We would like to thank Jan Witowski for assistance with running some of the models. We would like to thank Sarah Eskreiss Winkler for pointers to some of these models and data. This work was supported by NIH grants R01CA247910 and U54CA132378.

**REFERENCES**

1. U.S. Breast Cancer Statistics. *Breastcancer.org*

   https://www.breastcancer.org/symptoms/understand_bc/statistics (2021).

2. Berry, D. A. *et al.* Effect of Screening and Adjuvant Therapy on Mortality from Breast Cancer. *N. Engl. J. Med.* **353**, 1784–1792 (2005).

3. Screening Mammography | Health First Breast Center.

   https://hf.org/breasthealth/digitalmammo.cfm.

4. Lehman, C. D. *et al.* Diagnostic Accuracy of Digital Screening Mammography With and Without Computer-Aided Detection. *JAMA Intern. Med.* **175**, 1828–1837 (2015).

5. Geras, K. J. *et al.* High-Resolution Breast Cancer Screening with Multi-View Deep Convolutional Neural Networks. *ArXiv170307047 Cs Stat* (2018).

6. Lotter, W., Sorensen, G. & Cox, D. A Multi-Scale CNN and Curriculum Learning Strategy for Mammogram Classification. *ArXiv170706978 Cs* (2017).

7. Yala, A., Schuster, T., Miles, R., Barzilay, R. & Lehman, C. A Deep Learning Model to Triage Screening Mammograms: A Simulation Study. *Radiology* **293**, 38–46 (2019).

8. Ribli, D., Horváth, A., Unger, Z., Pollner, P. & Csabai, I. Detecting and classifying lesions in mammograms with Deep Learning. *Sci. Rep.* **8**, 4165 (2018).

9. Fatima, N., Liu, L., Hong, S. & Ahmed, H. Prediction of Breast Cancer, Comparative Review of Machine Learning Techniques, and Their Analysis. *IEEE Access* **8**, 150360–150376 (2020).

10. Shen, L. *et al.* Deep Learning to Improve Breast Cancer Detection on Screening Mammography. *Sci. Rep.* **9**, 12495 (2019).

11. McKinney, S. M. *et al.* International evaluation of an AI system for breast cancer screening. *Nature* **577**, 89–94 (2020).

12. Lotter, W. *et al.* Robust breast cancer detection in mammography and digital breast tomosynthesis using an annotation-efficient deep learning approach. *Nat. Med.* **27**, 244–249 (2021).

13. Shen, Y. *et al.* An interpretable classifier for high-resolution breast cancer screening

images utilizing weakly supervised localization. *Med. Image Anal.* **68**, 101908 (2021).

14. Yala, A. *et al.* Toward robust mammography-based models for breast cancer risk. *Sci. Transl. Med.* **13**, eaba4373 (2021).

15. Wu, N. *et al.* Deep Neural Networks Improve Radiologists' Performance in Breast Cancer Screening. *IEEE Trans. Med. Imaging* **39**, 1184–1194 (2020).

16. Liu, K. *et al.* Weakly-supervised High-resolution Segmentation of Mammography Images for Breast Cancer Diagnosis. *ArXiv210607049 Cs* (2021).

17. Moreira, I. C. *et al.* INbreast: toward a full-field digital mammographic database. *Acad. Radiol.* **19**, 236–248 (2012).

18. Lee, R. S. *et al.* A curated mammography data set for use in computer-aided detection and diagnosis research. *Sci. Data* **4**, 170177 (2017).

19. Cui, C. *et al.* The Chinese Mammography Database (CMMD): An online mammography database with biopsy confirmed types for machine diagnosis of breast. (2021) doi:10.7937/TCIA.EQDE-4B16.

20. Stadnick, B. *et al.* Meta-repository of screening mammography classifiers. (2022).

21. Suh, Y. J., Jung, J. & Cho, B.-J. Automated Breast Cancer Detection in Digital Mammograms of Various Densities via Deep Learning. *J. Pers. Med.* **10**, E211 (2020).

22. Kumari, L. & Jagadesh, B. A Robust Feature Extraction Technique for Breast Cancer Detection using Digital Mammograms based on Advanced GLCM Approach. *EAI Endorsed Trans. Pervasive Health Technol.* **8**, 172813 (2022).

23. Abdelhafiz, D., Yang, C., Ammar, R. & Nabavi, S. Deep convolutional neural networks for mammography: advances, challenges and applications. *BMC Bioinformatics* **20**, 281 (2019).

24. Chan, H.-P., Samala, R. K. & Hadjiiski, L. M. CAD and AI for breast cancer—recent development and challenges. *Br. J. Radiol.* **93**, 20190580 (2020).

25. Gardezi, S. J. S., Elazab, A., Lei, B. & Wang, T. Breast Cancer Detection and Diagnosis Using Mammographic Data: Systematic Review. *J. Med. Internet Res.* **21**, e14464 (2019).

26. CBIS-DDSM - The Cancer Imaging Archive (TCIA) Public Access - Cancer Imaging

Archive Wiki. https://wiki.cancerimagingarchive.net/display/Public/CBIS-DDSM.

27. INbreast Dataset | Kaggle. https://www.kaggle.com/ramanathansp20/inbreast-dataset.

28. Clark, K. *et al.* The Cancer Imaging Archive (TCIA): Maintaining and Operating a Public Information Repository. *J. Digit. Imaging* **26**, 1045–1057 (2013).

29. Huff, T., Mahabadi, N. & Tadi, P. Neuroanatomy, Visual Cortex. in *StatPearls* (StatPearls Publishing, 2022).